# Theoretical Estimation of Attenuation Coefficient of Resonant Ultrasound Contrast Agents

Lang Xia

Email: lang4xia@gmail.com

**Abstract**

Acoustic characterization of ultrasound contrast agents (UCAs, coated microbubbles) relies on the attenuation theory that assumes the UCAs oscillate linearly at sufficiently low excitation pressures. Effective shell parameters of the UCAs can be estimated by fitting a theoretical attenuation curve to experimentally measured attenuation data. Depending on the excitation frequency and properties of the shell, however, an UCA may oscillate nonlinearly even at sufficiently low excitation pressures violating the assumption in the linear attenuation theory. Notably, the concern on the estimation of the attenuation coefficient of a microbubble at resonance using linearized approximation has long been addressed. In this article, we investigated the attenuation phenomenon through analyzing the energy dissipation of a single UCA and propagating waves in an UCA suspension, both of which employed a nonlinear Rayleigh-Plesset equation. Analytical formulae capable of estimating the attenuation coefficient due to the weakly nonlinear oscillations of the UCA were obtained with a relatively rigorous mathematical analysis. The computed results that were verified by numerical simulations showed the attenuation coefficient of the UCA at resonance was pressure-dependent and could be significantly smaller than that predicted by the linear attenuation theory. Polydispersity of the UCA population enlarged the difference in the estimation of attenuation between the linear and present second-order nonlinear theories.

**I. INTRODUCTION**

Ultrasound Contrast Agents (UCAs), which are microbubbles coated with lipids, proteins, and polymers, have been studied in the medical field for diagnostic and therapeutic purposes due to their rich dynamic behaviors under the excitation of ultrasound (de Jong et al., 2000; Doinikov, Haac, & Dayton, 2009; Hoff, Sontum, & Hovem, 2000; Karandish et al., 2018; Kulkarni et al., 2018; Xia et al., 2017). The physical properties of the shell of an UCA play a key role in understanding the oscillations of the UCA in an acoustic field. Various shell models, conjugated with Rayleigh-Plesset type equations, have been proposed to characterize encapsulations of UCAs with or without thickness (Chatterjee & Sarkar, 2003; Church, 1995; de Jong, Hoff, Skotland, & Bom, 1992; Marmottant et al., 2005). Detailed discussions on all the current shell models are found in two comprehensive review papers (Doinikov & Bouakaz, 2011; Faez et al., 2013). Material characterization of the shell of an UCA can be performed by examining the dynamic behaviors of the UCA under the excitation of ultrasound with the assistance of tools such as fast framing optical cameras (De Jong, Emmer, Van Wamel, & Versluis, 2009), flow cytometry (Tu et al., 2011), and acoustic bulk measurements (Hoff et al., 2000). Due to the simplicity and effectiveness, the acoustic measurement is still a favorable method in studying the shell parameters. In this method, the comparison between experimental attenuation data and a theoretical attenuation curve is conducted to estimate the shell parameters of an UCA, such as shell dilatational viscosity and elasticity (or shear and elastic moduli). The frequency-dependent attenuation data can be acquired by an ultrasonic transducer operated in pulse-echo mode with low excitation pressures. The theoretical attenuation curve is obtained through solving a linearized Rayleigh-Plesset equation with a shell model, of which the attenuation mechanism is analog to that of a driven harmonic oscillator. Thus the energy dissipation (damping) at the linear region is estimated and used to calculate acoustic attenuation coefficient by the summation of the energy loss of each UCA (extinction cross-section) in the unit volume (de Jong et al., 1992; Medwin, 1977). Shell parameters then can be inversely solved by fitting the theoretical attenuation curve to the experimental attenuation data (Hoff et al., 2000; Paul et al., 2010). Yet another method, assuming the suspension of UCAs an effective medium (a bubbly liquid) and utilizing a linearized Rayleigh-Plesset equation, computes the attenuation of propagating acoustic waves in the medium by means of a dispersion relation derived from an effective wave equation (Commander & Prosperetti, 1989). These two methods produce identical results at low bubble volume fractions (Hoff, 2001; Xia, 2018). In both of the above methods, the explicity and validity of the linear attenuation formula (Medwin's) is crucial in solving the inverse problem (a fitting process).

It is well known that oscillation amplitudes of a gas bubble will be amplified significantly if the excitation frequency of impinging ultrasound closes to the resonance (or the main resonance) or harmonic resonance (the second resonance) frequencies. The linear acoustic properties of the bubble can be altered when it oscillates at resonance frequencies. Hilgenfeldt *et al.* (1998) analyzed the backscattering and absorption cross-sections of a resonant free bubble with an analogy to a nonlinear oscillator (Hilgenfeldt, Lohse, & Zomack, 1998). Krismatullin (2004) studied the effect of viscous damping on the resonance frequency of an encapsulated microbubble (Khismatullin, 2004). Ainslie and Leighton (2009) discussed the radiation damping of resonant bubbles using a linear equation of harmonic oscillator (Ainslie & Leighton, 2009). This result was then derived more systematically by using a Keller-type equation (Zhang, 2013).

Furthermore, the resonant oscillation of a bubble could induce thermal effect (Prosperetti, 1977), shape oscillation (Feng & Leal, 1993; Versluis et al., 2010), and the nonmonotonical and nonlinear oscillations (Lauterborn & Kurz, 2010). These nonlinear phenomena conspire to bring enormous complexities in the accurate estimation of the attenuation coefficient of UCAs that are based on encapsulated microbubbles. Currently, the thermal effect in the attenuation estimation may be dealt with by applying an isothermal process providing that all sizes of standard bubbles exceed 30 nm (Ainslie & Leighton, 2011). Additionally, for UCAs with diameters of a few micrometers, the thermal damping is two orders lower than the shell damping (Hoff et al., 2000). For small bubbles excited by a short pulse (or broadband in frequency domain due to the duality of Fourier transform), shape oscillation is not likely to happen, particularly in the initial few cycles (Versluis et al., 2010). To correct the linear attenuation at resonance, researchers have introduced radiation damping to account for the liquid compressibility due to vigorous oscillations of a bubble or UCA (Ainslie & Leighton, 2009; Paul et al., 2010; Xia, Porter, & Sarkar, 2015). While numerous experiments, simulations, and theories have addressed the concern on the attenuation estimation at resonance frequency of an UCA, an explicit attenuation formula, adding contribution due to nonlinear or large-amplitude oscillations to the Medwin's result (Medwin, 1977), has barely been found in the literature.

Therefore, in this article, by ignoring the nonlinear shell behaviors of an UCA, we study the attenuation due to nonlinear oscillations of the UCA via a modified Rayleigh-Plesset equation. An explicit attenuation formula, considering the contributions of weakly nonlinear dynamics of the UCA and the compressibility of the host liquid, is presented. The proposed formula is capable of dealing with attenuation due to nonlinear oscillations at resonance, of which the results at low excitation pressures agrees with the Medwin's classic result. We also employ an approach of effective medium to derive a nonlinear attenuation formula of propagating waves in the UCA suspension to verify the results calculated by the proposed formula. Finally, we discuss the effects of higher-order nonlinearities, excitation pressures, and size distribution on the attenuation phenomenon due to resonant oscillations of the UCA.

## II. MATHEMATICAL ANALYSIS

We briefly present the development of linear attenuation theory due to bubble oscillations, as well as perturbative analysis performed on bubble oscillation and wave propagation in the present section.

### A. Bubble dynamics

By neglecting gas diffusion and assuming spherical pulsation, a modified form of Rayleigh-Plesset equation for an UCA coated with a viscoelastic shell oscillating in a compressible liquid can be written in the form of (Marmottant et al., 2005)

$$\rho\left(R\ddot{R}+\frac{3}{2}\dot{R}^2\right)=P_{g0}\left(\frac{R_0}{R}\right)^{3\kappa}\left(1-\frac{3\kappa\dot{R}}{c_0}\right)-4\kappa^s\frac{\dot{R}}{R^2}-2E^s\frac{1}{R}\left(\frac{R^2}{R_0^2}-1\right)-2\gamma_0\frac{1}{R}-4\mu\frac{\dot{R}}{R}-p_0+p(t) \quad (1)$$

where $\rho$ is the density of the surrounding liquid, $R$ is the instantaneous radius of the bubble around the equilibrium radius $R_0$, and $\dot{R}=dR/dt$, $\ddot{R}=d^2R/dt^2$, $P_{g0}$ is the initial pressure inside the bubble, $\kappa$ is the polytropic constant, $c_0$ is the sound speed in the host liquid without UCAs, $\kappa^s$ and $E^s$ are the dilatational viscosity and elasticity of the UCA shell, respectively, $\mu$ is the viscosity of the host liquid, $\gamma_0$ is a reference value of the surface tension, $p_0$ is the hydrostatic pressure in the liquid, and $p(t)$ is the excitation pressure. Note that in the above equation we have assumed the elasticity to be a constant in lieu of a fractional function accounting for a bulking effect of the bubble shell that is shown in the original equation. Although different types of Rayleigh-Plesset equations exist due to various shell material, the above equation is a popular modification and well accepted for modeling the dynamics of lipid-coated UCAs. Interested readers are referred to this comprehensive review (Doinikov & Bouakaz, 2011). Equation (1) can be transformed into a constant elasticity model (CEM) by replacing the initial radius with the equilibrium radius of the UCA (Sarkar, Shi, Chatterjee, & Forsberg, 2005).

## B. Small-amplitude approximation

The dynamical equation for the UCA of equilibrium radius $R_0$, undergoing forced spherical pulsation [ $R(t)=R_0+X(t)$ and $|X(t)|<<R_0$ ] up till to the second-order approximation at an external excitation pressure $p(t)$ can be written in the form of

$$\left(\ddot{X}+\omega_1\delta_1\dot{X}+\omega_1^2 X\right)+\frac{X}{R_0}\left(\ddot{X}-(\omega_2\delta_2-\frac{3}{2}\frac{\dot{X}}{X})\dot{X}-\omega_2^2 X\right)=\frac{p(t)}{\rho R_0} \quad (2)$$

in which the following definitions have been adopted

$$\omega_1^2 = \frac{1}{\rho R_0^2}\left(3\kappa p_0 + (3\kappa-1)\frac{2\gamma_0}{R_0} + \frac{4E^s}{R_0}\right)$$

$$\delta_1 = \frac{4}{\rho\omega_1 R_0^2}\left(\frac{\kappa^s}{R_0} + \frac{3\kappa R_0 P_{g0}}{4c_0} + \mu\right) \tag{3}$$

and

$$\omega_2^2 = \frac{1}{2\rho R_0^2}\left(3\kappa(3\kappa+1)p_0 + (3\kappa+2)(3\kappa-1)\frac{2\gamma_0}{R_0} + \frac{4E^s}{R_0}\right)$$

$$\delta_2 = \frac{4}{\rho\omega_2 R_0^2}\left(\frac{2\kappa^s}{R_0} + \frac{9\kappa^2 R_0 P_{g0}}{4c_0} + \mu\right) \tag{4}$$

Equation (2) can be used to model weakly nonlinear oscillations of the UCA at relatively higher excitation pressures.

## C. Linear theory

By neglecting the second-order term in Eq.(2), the dynamical equation for the UCA undergoing forced linear spherical oscillation at an external excitation pressure $p(t) = P_A \cos(\omega t)$ can be reduced into the following equation of a harmonic oscillator

$$\ddot{X}_l + \omega_1 \delta_1 \dot{X}_l + \omega_1^2 X_l = \frac{P_A}{\rho R_0}\cos(\omega t) \tag{5}$$

with the damping constant defined by $\delta_1 = \delta_{shell} + \delta_{liquid} + \delta_{radiation}$ and

$$\delta_{shell} = \frac{4\kappa^s}{\rho\omega_1 R_0^3}, \quad \delta_{liquid} = \frac{4\mu}{\rho\omega_1 R_0^2}, \quad \delta_{radiation} = \frac{1}{\rho\omega_1 R_0}\frac{3\kappa P_{g0}}{c_0} \tag{6}$$

where $X_l$ is the first-order displacement around equilibrium radius $R_0$, $\omega_1$ is the UCA's pulsation angular natural frequency, which are used as the UCA's resonance frequency in this article. The radiation damping $\delta_{radiation}$ is due to the compressibility of the host medium presumably accounting for large-amplitude oscillations of the UCA at resonance. Eq.(5) indicates that the linear dynamics of a coated microbubble (UCA) is a linear harmonic oscillator, having a steady-state solution in the form of

$$X_l = \frac{1}{2}\left[A\exp(i\omega t) + \bar{A}\exp(-i\omega t)\right]$$
$$A = \frac{P_A}{\rho R_0} \frac{1}{-\omega^2 + i\omega\omega_1\delta_1 + \omega_1^2} \quad (7)$$

Here $\bar{A}$ is the complex conjugate of $A$.

## III. ESTIMATION OF LINEAR ATTENUATION

### A. Method of harmonic-oscillator

To estimate the oscillation induced attenuation, we can investigate the energy absorption of a bubble in the acoustic field (Xia, 2018). By assuming that an UCA oscillates spherically without thermal dissipation in a liquid, the energy delivered to the UCA per unit time can be written as

$$\Pi_i = p\frac{d}{dt}\left(\frac{4}{3}\pi R^3\right) = (4\pi R^2 p)\dot{R} \quad (8)$$

The average power delivered into the system is

$$\Pi = \frac{1}{T}\int_0^T \Pi_i dt \quad (9)$$

where $T = 2\pi/\omega$ is the period. The average intensity of the impinging acoustic wave can be written as $I = P_A^2/(2\rho c_0)$. Here $c_0$ is the sound speed in the host liquid, and the extinction cross-section is given by

$$\sigma_e = \frac{\Pi}{I} = \frac{4\omega\rho c_0}{P_A}\int_0^{2\pi/\omega} R^2\dot{R}\cos(\omega t)dt \quad (10)$$

The above equation is the extinction cross-section of a microbubble subjected to acoustic excitation. Here we only investigate the effects of bubble oscillation on the energy attenuation, detail discussions on other mechanisms of dissipation may refer to a comprehensive review (Ainslie & Leighton, 2011). By substituting the real part of the solution Eq.(7) into Eq.(10), the linear extinction cross-section $\sigma_1$ is given by the following equation

$$\sigma_1 = 4\pi R_0^2 \rho c\omega \frac{\text{Im}\{-A\}}{P_A} \quad (11)$$

or in the form of (Medwin, 1977)

$$\sigma_1 = 4\pi R_0^2 \frac{\delta_1 c_0}{\omega_1 R_0} \frac{\Omega^2}{\left(1-\Omega^2\right)^2 + \delta_1^2 \Omega^2} \tag{12}$$

where $\Omega = \omega/\omega_1$. The attenuation coefficient $\alpha$ of a bubble suspension can be computed readily by assuming a linear attenuation law (de Jong et al., 1992), of which the final result is given by

$$\alpha = (20\log e)\frac{1}{2}\int_0^\infty \sigma_1(\omega, R_0)n(R_0)dR_0 \quad [\text{dB/m}] \tag{13}$$

where $e$ is the base of the natural logarithm, and $n$ is the bubble size distribution. It is not difficult to see that the attenuation coefficient for an UCA suspension of monodisperse population is proportional to the extinction cross-section of the individual UCA.

**B. Method of effective-medium**

When bubble population in the host liquid is small such that bubble-bubble interactions and multiple scatterings are negligible, the suspension of the bubbles can be treated as an effective homogeneous medium, which is called a bubbly liquid. By applying the continuum and momentum equations, one can obtain the following equations characterizing acoustic waves propagating in bubbly liquids (Commander & Prosperetti, 1989)

$$\frac{1}{c_0^2}\frac{\partial^2 p}{\partial t^2} - \frac{\partial^2 p}{\partial x^2} = \rho\frac{\partial^2 \beta}{\partial t^2}, \quad \text{and} \quad \beta = \frac{4}{3}\pi\int_0^\infty R^3 n(R_0)dR_0 \tag{14}$$

Here $\beta$ is the bubble volume fraction. The instantaneous radius $R$ in the above equation is also the function of the spatial variable $x$. By substituting Eq.(7) (with the spatial variable being incorporated) into Eq.(14) and neglecting the second and higher-order terms, one can obtain a dispersion relation

$$k_1^2 = \frac{\omega^2}{c_0^2} - 4\pi\int_0^{+\infty} \frac{\omega^2}{-\omega^2 + i\omega\omega_1\delta_1 + \omega_1^2}R_0 n(R_0)dR_0 \tag{15}$$

where $k_1$ is a complex wave number. Let $k_1 = k_{1R} - i\alpha_1$, then the attenuation coefficient is identified by

$$\alpha = 20(\log e)\alpha_1 \text{ dB/m} \tag{16}$$

An underlying assumption in Eq.(15) is that the bubble volume fraction $\beta$ is very small, which should be below $10^{-4}$ (Xia, 2019). If the concentration of UCAs in the suspension is small, one can perform Taylor expansion with respect to $\beta$ and keep only the first-order term. Then it can be easily proved that Eq.(16) and Eq.(13) predict the same results (Xia, 2018). Note that the concentration of microbubbles in UCA suspensions is usually much lower than this value ($10^{-4}$), we can safely use the method of the harmonic oscillator in the linear analysis as a matter of simplicity and convenience.

## IV. ATTENUATION DUE TO NONLINEAR OSCILLATIONS

The attenuation coefficient of an UCA estimated by Eq.(12), which only the first-order solution Eq.(7) is considered, is valid for the UCA oscillating at sufficient small amplitudes. This suggests the formula is insufficient in the estimation of attenuation coefficient due to nonlinear or large-amplitude oscillations of the UCA at resonance frequencies. Full solutions of Eq.(1) are needed to investigate the nonlinear attenuation. Note that the analytical solutions of Eq.(1) are not available by far, one may numerically solve the equation and feed the solved radius-time data into Eq.(10) to calculate the frequency-dependent attenuation curve. However, an analytical expression of the attenuation coefficient is still crucial in the material characterization of UCAs for obtaining the shell parameters. Currently, effective and explicit formulas for calculating the nonlinear attenuation at resonance frequencies of UCAs are still under development. We thus try to obtain an analytical expression for the nonlinear attenuation coefficient by employing the following perturbative approximation

$$X = X_l + Y \tag{17}$$

Here $Y$ is the nonlinear part of the oscillation. Substitution of the above equation into Eq.(2) yields

$$\ddot{Y} + \omega_1 \delta_1 \dot{Y} + \omega_1^2 Y = f(X_l, Y) \tag{18}$$

where

$$f(X_l, Y) = -\frac{3}{2R_0}(\dot{X}_l + \dot{Y})^2 - \frac{1}{R_0}(X_l + Y)\left[(\ddot{X}_l + \ddot{Y}) - \omega_2 \delta_2 (\dot{X}_l + \dot{Y}) - \frac{1}{2}\omega_2^2 (X_l + Y)\right] \tag{19}$$

Although $Y$ in the left-hand side is the same as that of the right-hand side, $f(X_l, Y)$ may be considered as a driving force. In this sense, only the frequency or their combinations that are closed to $\omega_1$ (e.g., the resonance) will have a significant impact on the dynamics of $Y$ (Prosperetti, 2013). Therefore, in the second harmonic region, we may write a steady solution in the form of

$$Y = \frac{1}{2}\left[B\exp(2i\omega t) + \bar{B}\exp(-2i\omega t)\right] \tag{20}$$

Again $\bar{B}$ is the complex conjugate of $B$. By substituting the above equation into (18) and neglecting the terms and their combinations that cannot give rise to second harmonics, we finally have the solution in the form of

$$X = \frac{1}{2}\left[A\exp(i\omega t) + \bar{A}\exp(-i\omega t)\right] + \frac{1}{2}\left[B\exp(2i\omega t) + \bar{B}\exp(-2i\omega t)\right] \tag{21}$$

or

$$X = \frac{1}{2}\left[A\exp(i\omega t) + B\exp(2i\omega t)\right] + c.c. \tag{22}$$

where $c.c.$ shorts for the complex conjugate of the preceding terms, and

$$B = \frac{A^2}{4R_0} \frac{5\omega^2 + 2i\delta_2\omega_2\omega + 2\omega_2^2}{-4\omega^2 + 2i\delta_1\omega_1\omega + \omega_1^2} \tag{23}$$

Here the coefficients of the above equation have been expressed in Eq.(3) and Eq.(4). By the substitution of Eq.(22) into Eq.(10) and neglecting all terms that do not give rise to the fundamental and second harmonic components, we obtain the equation for estimating the attenuation coefficient of UCAs at nonlinear oscillations

$$\sigma_e = \left(1 - \frac{\text{Re}\{B\}}{R_0}\right)\sigma_1 + \frac{\text{Re}\{A\}}{R_0}\sigma_2 \tag{24}$$

where

$$\sigma_2 = 4\pi R_0^2 \rho c \omega \frac{\text{Im}\{-B\}}{P_A} \tag{25}$$

Equation (24), which is the major result in this article, indicates that the nonlinear parts are introduced in the attenuation of large amplitude oscillations. It reduces to the Medwin's formula (12) when the in-phase amplitude $\text{Re}\{A\}, \text{Re}\{B\} \ll R_0$. Detail discussion on this result will be presented in the Result section.

**IV. ATTENUATION DUE TO NONLINEAR PROPAGATION**

In the preceding section, we employed the method of harmonic oscillator and investigated the attenuation due to the nonlinear oscillations of the UCA, with an underlying assumption that the oscillating UCA does not distort the impinging acoustic waves. While the assumption is legitimate in linear theory, its validity in the nonlinear theory needs further discussion. We thus employ the method of effective medium to study the effect of nonlinear oscillations of UCAs on the propagating acoustic waves. Note that the flow field in the medium induced by the propagating waves cannot be investigated directly via Eq.(14), we don't perform the perturbative analysis on the velocity field, which may contribute to the attenuation due to the second order effect, e.g., acoustic streaming. We follow the same perturbation analysis by introducing additional spatial variables and write

$$p(x,t) = \frac{1}{2}\left[p_1(x)\exp(i\omega t) + p_2(x)\exp(2i\omega t)\right] + c.c.$$
$$X(x,t) = \frac{1}{2}\left[A(x)\exp(i\omega t) + B(x)\exp(2i\omega t)\right] + c.c.$$
(26)

Here the $p_1(x)$ and $p_2(x)$ are the first and second-order approximations, respectively. Eq.(26) indicates that the external periodic force $p(t)$ in Eq.(2) becomes the propagating waves $p(x,t)$ in the UCA suspension, and $X$ is now a function of both time and space. Accordingly, we can obtain the expressions for $A(x)$ and $B(x)$ in terms of Eq.(5) and Eq.(19)

$$A(x) = \frac{1}{-\omega^2 + i\omega\omega_1\delta_1 + \omega_1^2}\frac{p_1(x)}{\rho R_0}$$
$$B(x) = \frac{5\omega^2 + 2i\delta_2\omega_2\omega + 2\omega_2^2}{-4\omega^2 + 2i\delta_1\omega_1\omega + \omega_1^2}\frac{A(x)^2}{4R_0} + \frac{1}{-4\omega^2 + 2i\delta_1\omega_1\omega + \omega_1^2}\frac{p_2(x)}{\rho R_0}$$
(27)

$B(x)$ in the above equation has contributions from the second-order acoustic waves, which is neglected in Eq.(23). By the substitution of Eq.(26) into Eq.(14), we can obtain the solution for the first and second or the pressure terms

$$p_1(x) = P_A \exp(-ik_1 x)$$
$$p_2(x) = \frac{2\Gamma\omega^2 P_A^2}{\rho c_0^4\left(4k_1^2 - k_2^2\right)}\left[\exp(-ik_2 x) - \exp(-2ik_1 x)\right]$$
(28)

where

$$\Gamma = 2\pi \int_0^{+\infty} \frac{c_0^4}{R_0\left(-\omega^2 + i\omega\omega_1\delta_1 + \omega_1^2\right)^2}\left(2 + \frac{5\omega^2 + 2i\delta_2\omega_2\omega + 2\omega_2^2}{-4\omega^2 + 2i\delta_1\omega_1\omega + \omega_1^2}\right)n(R_0)dR_0$$

$$k_2^2 = 4\omega^2\left(\frac{1}{c_0^2} - 4\pi\int_0^{+\infty}\frac{1}{-4\omega^2 + 2i\delta_1\omega_1\omega + \omega_1^2}R_0 n(R_0)dR_0\right)$$

(29)

$\Gamma$ in the above equation is related to the nonlinear coefficient of the medium (Xia, 2019). $k_1$ and $k_2$ are the complex wavenumbers of the acoustic waves at fundamental and second harmonic frequencies, respectively. Finally, we can obtain the solution of the wave equation in the form of

$$p(x,t) = \frac{1}{2}P_A \exp(i\omega t - ik_1 x) + \frac{\Gamma\omega^2 P_A^2}{\rho c_0^4\left(4k_1^2 - k_2^2\right)}\left[\exp(2i\omega t - ik_2 x) - \exp(2i\omega t - 2ik_1 x)\right] + c.c. \quad (30)$$

and it is rearranged in the following form

$$p(x,t) = \frac{1}{2}p_1(x,t) + \frac{\Gamma\omega^2}{\rho c_0^4\left(4k_1^2 - k_2^2\right)}\left[\exp(\Delta\alpha x)\exp(i\Delta k x) - 1\right]p_1^2(x,t) + c.c. \quad (31)$$

where $p_1(x,t) = P_A \exp(-\alpha_1 x)\exp(i\omega t - ik_{1R}x)$, $k_2 = k_{2R} - i\alpha_2$, $\Delta\alpha = 2\alpha_1 - \alpha_2$, and $\Delta k = 2k_{1R} - k_{2R}$. The above equation shows that the nonlinear coefficient $\Gamma$, which is crucial in the generation of second harmonic waves, will bring additional difficulty in the evaluation of the attenuation coefficient of the medium as the attenuation is not a simple summation of the imaginary parts of $k_1$ and $k_2$. However, the attenuation of the propagating wave is characterized by the decreasing of the envelope of the wave $p$, which can be approximated by

$$\tilde{p}(x) = \frac{1}{2}P_A \exp(-\alpha_1 x) + \frac{\Gamma\omega^2}{\rho c_0^4\left(4k_1^2 - k_2^2\right)}\left[\exp(\Delta\alpha x)\exp(i\Delta k x) - 1\right]P_A^2 \exp(-2\alpha_1 x) + c.c. \quad (32)$$

Therefore, the attenuation coefficient can be determined by

$$\tilde{\alpha}(\omega) = -\frac{1}{x}\ln\frac{|\tilde{p}(x)|}{|\tilde{p}(0)|} \quad (33)$$

In order to estimate the attenuation, we may assume the second harmonic field is generated synchronous and accumulates with distance, then we may use $\Delta k = 2k_{1R} - k_{2R} = 0$ and obtain the attenuation coefficient in the form of

$$\alpha = \alpha_1 - \frac{1}{x}\ln\left[1+(P_B+\bar{P}_B)\exp(-\alpha_1 x)\right] \tag{34}$$

where the nondimensionalized amplitude at the second harmonic is defined by

$$P_B = \frac{\Gamma\omega^2}{\rho c_0^4\left(4k_1^2 - k_2^2\right)}\left[\exp(\Delta\alpha x)-1\right]P_A \tag{35}$$

Utilization of $x = 2\pi/|k_{1R}|$ gives a reduced form of the attenuation coefficient

$$\alpha = \alpha_1 - \frac{k_{1R}}{2\pi}\ln\left[1+(P_B+\bar{P}_B)\exp(-\alpha_1 \frac{2\pi}{k_{1R}})\right] \tag{36}$$

The above equation is the second primary result in this article. It suggests that the generation of the second harmonic wave decreases the magnitude of the attenuation coefficient at the resonance frequency, and the attenuation coefficient reduces to the classic linear result when the second harmonic amplitude is zero.

## IV. RESULTS AND DISCUSSIONS

### A. Numerical verification

To verify the accuracy of Eq.(24) obtained from perturbation analysis, we numerically solved the full Rayleigh-Plesset equation (1) in MATLAB with the internal ODE solver ODE45s. As for the verification of the analytical formula Eq.(36) in the model of effective medium, we solved the equation (1) and (14) simultaneously using a finite element solver in COMSOL. The physical constants of water (host medium) used in the numerical simulations are listed as follows: density $\rho = 1000\ \text{Kg/m}^3$, hydrostatic pressure $p_0 = 101325\ \text{Pa}$, sound speed $c_0 = 1500\ \text{m/s}$, viscosity $\mu = 0.002\ \text{N}\cdot\text{s/m}^2$ (is doubled to account for thermal effect). The physical constants of the UCA are: bubble radius $R_0 = 2.5\times 10^{-6}\ \text{m}$, polytropic constant $\kappa = 1.07$, shell viscosity of the $\kappa^s = 2.5\times 10^{-9}\ \text{N}\cdot\text{s/m}$, shell elasticity $E^s = 0.5\ \text{N/m}$, surface tension $\gamma_0 = 0$. Finally, we assumed the UCA population in the suspension to be monodisperse with the quantity $N = 10^{10}\ 1/\text{m}^3$.

### 1. Dynamics of a single UCA

Figure 1 lists the radius-frequency curves at four different excitation pressures of 0.1$p_0$ (a), 0.2 $p_0$ (b), 0.4 $p_0$ (c), and 0.8 $p_0$ (d). The selected excitation pressures are similar to those used in acoustic experiments for the measurement of acoustic attenuation. The '1st order, oscillator' is the analytical results calculated by Eq.(7), the '2nd order, oscillator' is the analytical results calculated by Eq.(22), and the 'numerical, oscillator' is the numerical results obtained from Eq.(1). At the lowest excitation pressure $0.1p_0$, it is not difficult to see from Figure 1a that all the solutions are closed to each other. The peak amplitude of the 1st order curve is slightly above the 2nd order and numerical results, while the latter two curves are identical. This suggests that the UCA oscillates almost linearly at the excitation pressure $0.1p_0$ near the resonance. Figure 1b shows that the peak amplitude of the 1st order curve deviates more from the other nonlinear results. Additionally, the 1st order curve does not predict a kink at the excitation frequency of $f/f_0 = 0.5$, which is the resonance of the second harmonic. The 2nd order curve still overlaps with the numerical curve at the excitation pressure of $0.2p_0$. When the excitation pressure is further increased to $0.4p_0$, Figure 1c presents enlarged differences between the linear and nonlinear results. Finally, in Figure 1d the differences among the three curves near the peaks can be easily differentiated. Near the main resonance where the excitation frequency is $f/f_0 = 1$, the linear result is about 30% larger than the nonlinear results. The 2nd order and numerical curves have almost the same peak value, however, the location of the peak in the numerical curve skews to the right located at $f/f_0 = 1.06$. The 2nd order and numerical curves are identical near the second resonance $f/f_0 = 0.5$. The above comparisons suggest that the 2nd order approximation may be considered as a more accurate solution than the linear solutions to model the weakly nonlinear resonant oscillations of the UCA.

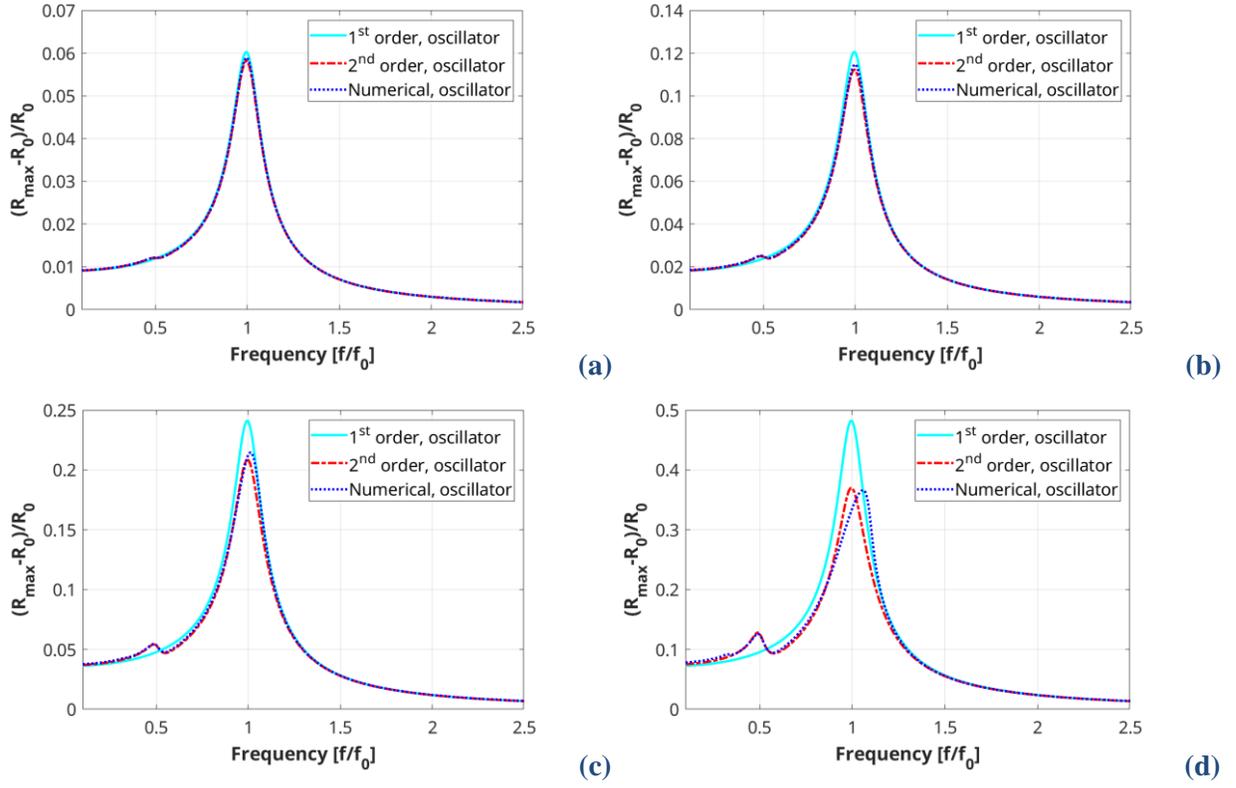

**Figure 1**: The normalized maximum bubble radius *vs.* frequency at the excitation pressure of $0.1p_0$ (a), $0.2p_0$ (b), $0.4p_0$ (c), and $0.8p_0$ (d).

## 2. Propagation of acoustic wave in the UCA suspension

In the effective medium model, the propagation waves in the suspension of monodisperse UCAs are compared among linear, nonlinear, and numerical results. Figure 2 displays the last step of the propagation waves at the frequency of the UCA resonance ($f/f_0 = 1$). The linear results, as denoted by '1$^{st}$ order, wave', are calculated by the first equation in Eq.(26) by assuming $p_2(x) = 0$. The nonlinear analytical results calculated by Eq.(30) are denoted by '2$^{nd}$ order, wave', and the nonlinear numerical results obtained from Eq.(14) are denoted by 'numerical, wave'. The curves in each of the figures show a great similarity at the excitation pressures up till $0.4p_0$, indicating the analytical results have a good approximation to the numerical solutions of the full wave equations. When the excitation pressure reaches $0.8p_0$, the distortion increases, as shown by the insert in Figure 2d, and both the 1$^{st}$ and 2$^{nd}$ order curves are different from the numerical curve (indicated by the blue dashed curve). However, the nonlinear analytical result '2$^{nd}$ order, wave' closely resembles the shape of the numerical curve. Thus, Eq.(30) is seen as a more accurate approximation than the linear result for capturing the nonlinear characteristic of propagating waves in the UCA suspension.

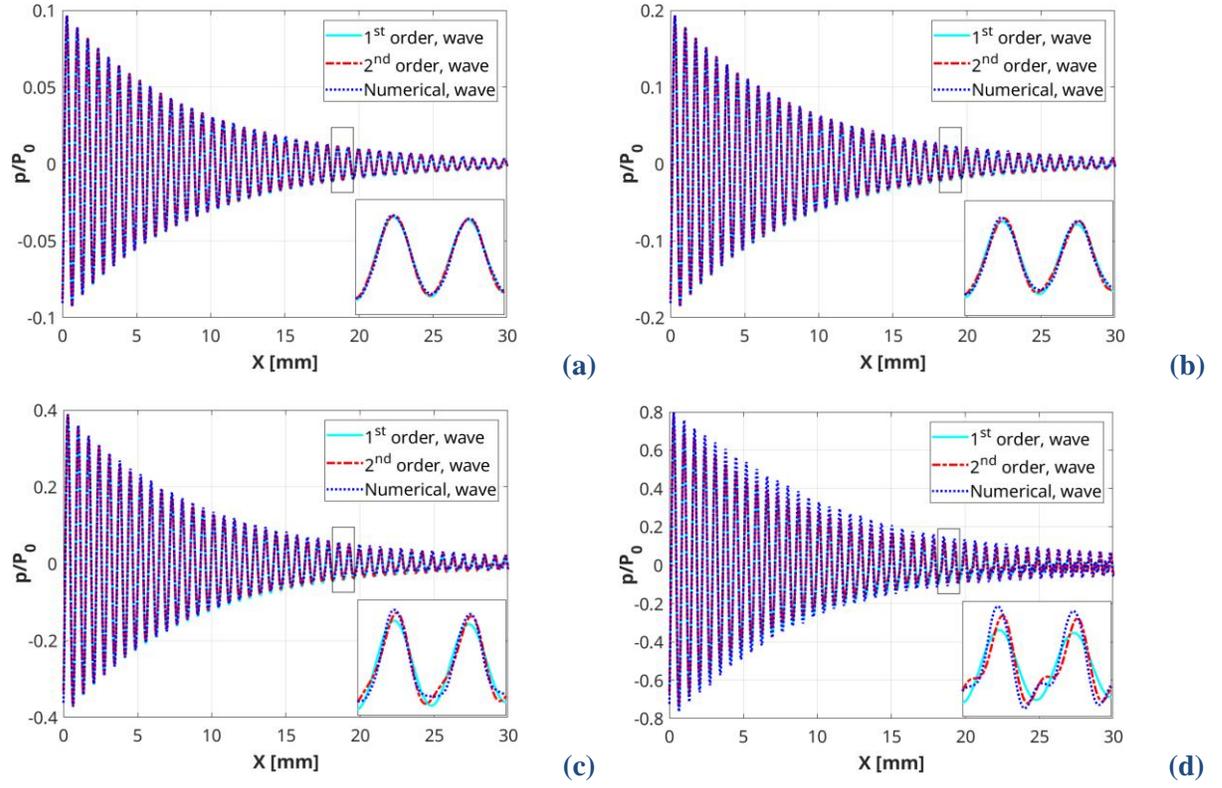

**Figure 2**: The profiles of propagation waves in the suspension of UCAs at the excitation pressure of $0.1p_0$ (a), $0.2p_0$ (b), $0.4p_0$ (c), and $0.8p_0$ (d). The insert in each figure is the zoomed-in view of the rectangle region. The frequency of the waves is the resonance frequency $f_0$ of the UCA.

**B. Comparisons between estimated attenuation coefficient**

The plots of attenuation coefficient derived from different solutions are presented in Figure 3. At the excitation pressure of $0.1p_0$, Figure 3a shows that all the curves almost overlap with each other and have the same peak value near the main resonance frequency of $f/f_0 = 1$, except that the result of the 2nd order effective medium [Eq.(36)] displays a small peak near the second resonance $f/f_0 = 0.5$. With increasing the excitation pressures, Figure 3b and c show increased differences between the linear attenuation (denoted by 'Medwin') and the nonlinear results (2nd order and numerical). The attenuation coefficients estimated by methods of oscillator and effective medium are almost identical near the main resonance, indicating the equivalence of the two methods for estimating attenuation near the main resonance. Note that the 2nd order effective medium overestimates the attenuation coefficient near the second resonance frequency. With further increasing the excitation pressure to $0.8p_0$, Figure 3d shows the separation of attenuation curves near the main resonance. The peak values of both 2nd order attenuation curves are closed to 20, which is about 13% less than that estimated by Medwin's formula. However, the attenuation curve obtained from the numerical solutions of the oscillator model is distorted, skewed to the right, and with the peak also locating at $f/f_0 = 1.06$, the same with the radius-frequency curve in Figure 1d. This result may raise questions on the accuracy of the 2nd order approximations for capturing nonlinear attenuation near the main resonance. However, an additional numerical result of the full wave equation from the method of effective medium, denoted by the circles, exhibits a close similarity to both of the 2nd order approximations near the main resonance. As for the attenuation at the second resonance, it gives a peak value of 1.41, which is larger than the value 0.34 of the '2nd order, oscillator', but much less than the value 4.9 of the '2nd order, wave'. Although near the second resonance frequency the attenuation estimation given by either the 2nd order oscillator or effective medium model provides a mediocre agreement to the numerical results of the effective medium, we still see that both the 2nd order approximations have great accuracy in the estimation of attenuation coefficient outside the second resonance region. Also, we may conclude that the methods of 2nd order oscillator and the method of 2nd order effective medium are equivalent for the estimation of the attenuation coefficient of a resonant UCA. Due to the simplicity of the results from the method of harmonic oscillator, Eq.(24) will be employed for the analysis in the rest of this article.

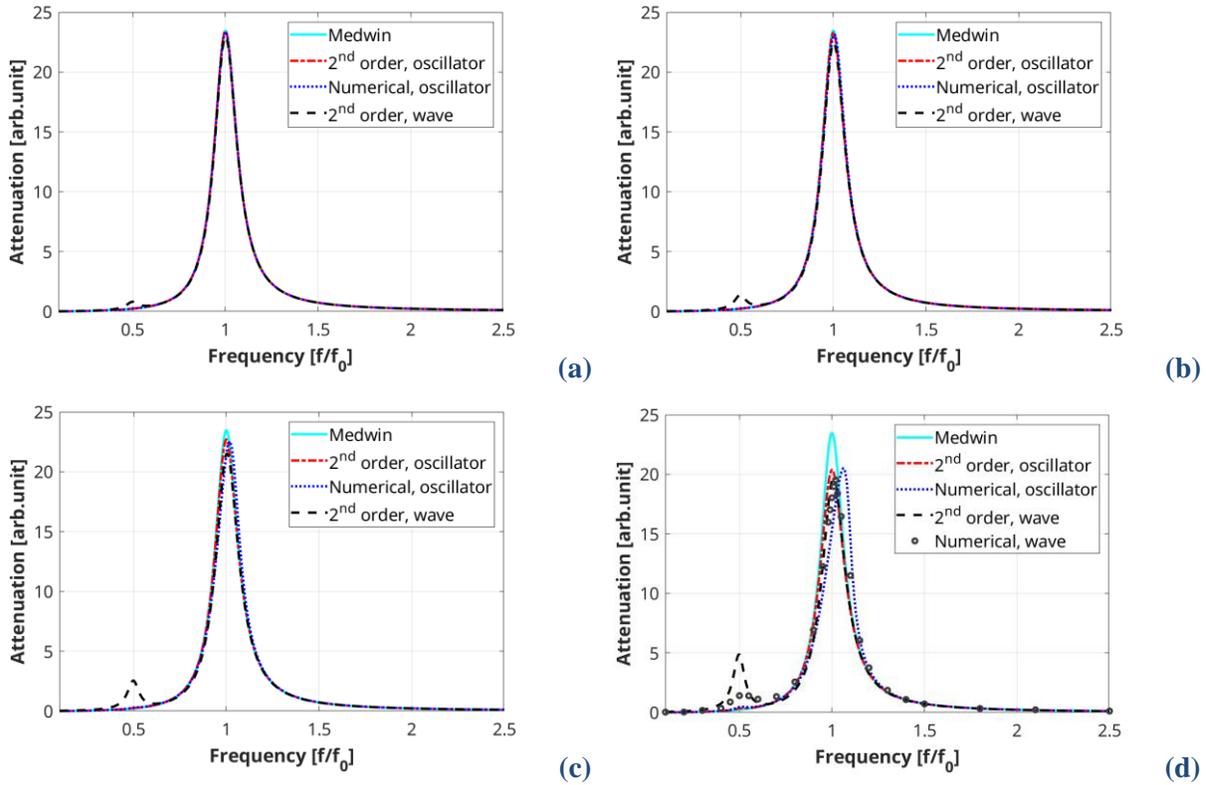

**Figure 3**: The attenuation coefficient *vs.* frequency at the excitation pressure of $0.1p_0$ (a), iterated computation with the excitation pressures being increased to $0.2\ p_0$ (b), $0.4\ p_0$ (c), and $0.8\ p_0$ (d).

## C. Comment on the higher-order nonlinearities

To understand the skewing of the main resonance peak near the fundamental frequency in the last section, we need approximate the RPE up till to the third-order approximation, which is in the form of

$$\left(\ddot{X} + \omega_1 \delta_1 \dot{X} + \omega_1^2 X\right) + \frac{X}{R_0}\left(\ddot{X} - (\omega_2 \delta_2 - \frac{3}{2}\frac{\dot{X}}{X})\dot{X} - \omega_2^2 X\right) + \left(\frac{X}{R_0}\right)^2 \left(\omega_3 \delta_3 \dot{X} + \omega_3^2 X\right) = \frac{p(t)}{\rho R_0} \quad (37)$$

in which the coefficient in the third-order term is defined by

$$\omega_3^2 = \frac{1}{2\rho R_0^2}\left(\kappa(3\kappa+1)(3\kappa+2)p_0 + (3\kappa^2+4\kappa+2)(3\kappa-1)\frac{2\gamma_0}{R_0} + \frac{4E^s}{R_0}\right)$$
$$\delta_3 = \frac{4}{\rho \omega_3 R_0^2}\left(\frac{3\kappa^s}{R_0} + \frac{9\kappa^2 R_0 P_{g0}}{8c_0} + \mu\right) \quad (38)$$

Using the same physical constant as in the preceding section, we display the comparison of the computed results in Figure 4. The radius curve of the 3rd order approximation is identical to the numerical result of the full equation at the excitation pressure of $0.8p_0$. This indicates that the skewing behavior of the resonance peak is caused by the third-order nonlinearity. In fact, the right shift of the peak refers to a 'hard nonlinearity'. Depending on the shell parameters, the resonance peak can also shift to the left (or 'soft nonlinearity'). Despite that the 2nd order approximation is not sufficient to represent the real radius curve of a single UCA, the corresponding attenuation curve (see Figure 3d) computed from the numerical solutions of the wave equation (14) does not have the skewing behavior induced by the third-order nonlinearity. This could be due to the dispersive nature of the UCA suspension counteracts the impacts of higher-order nonlinearities in the UCA suspension (Xia, 2019).

Therefore, for the range of excitation pressures employed in the present simulations, the 3rd order approximation does not offer better attenuation estimation than that of the 2nd order approximations for the UCA suspension. The proposed 2nd order formulae are accurate in the estimation of the attenuation coefficient near the main resonance of the UCA. Note that these employed excitation pressures are commonly used in the acoustic attenuation experiments for the material characterization of UCAs. The 3rd order nonlinear approximation would be necessary if the excitation pressures were further increased. However, this concern is out the scope of the present work, as higher excitation pressures are avoided in the attenuation experiments of UCA characterization.

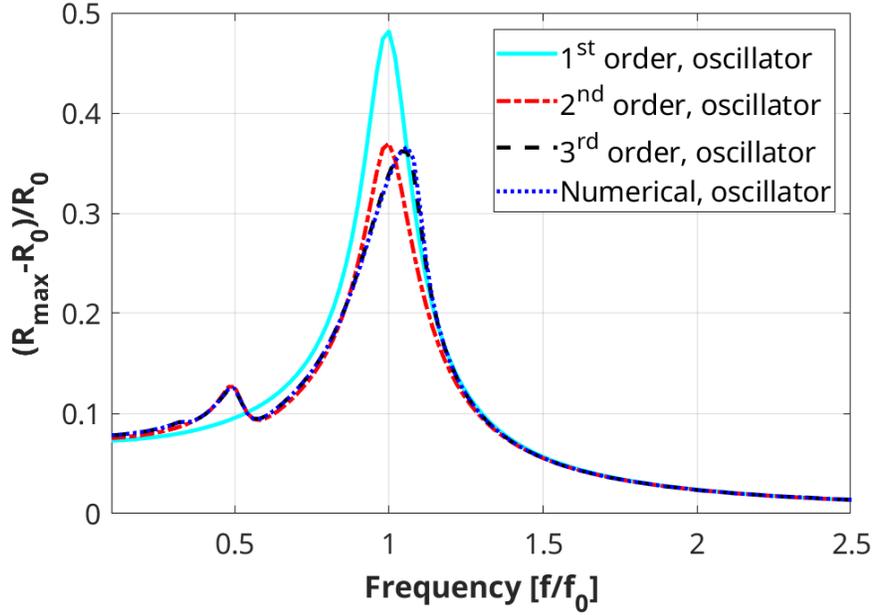

**Figure 4**: The normalized maximum bubble radius vs frequency at the excitation pressure of $0.8p_0$. The third order approximation (black-dashed curve) exactly matches the numerical result (blue short-dashed curve).

### D. Comment on the pressure-dependent attenuation

Nonlinear phenomena of UCA attenuation in acoustic experiments have been observed for decades, amongst which the pressure-dependent attenuation has been investigated by numerous authors (Chen, Zagzebski, Wilson, & Stiles, 2002; Emmer et al., 2009; Gong, Cabodi, & Porter, 2014; Helfield & Goertz, 2013; Overvelde et al., 2010; Segers, de Jong, & Versluis, 2016; Tang & Eckersley, 2007; Tang, Eckersley, & Noble, 2005; Xia et al., 2015). Those experimental observations cannot be easily interpreted in the framework of the linear attenuation theory (Medwin's formula). This is because, at sufficiently low excitation pressures, an UCA is usually assumed to oscillate linearly such that the attenuation coefficient estimated by the linear attenuation theory is independent of the excitation pressures. Most of the above experimental investigations on lipid-coated microbubbles reported that the peak attenuation coefficient near the main resonance increases with increasing the excitation pressures, whereas the present study demonstrates that the peak attenuation decreases with increasing the excitation pressure due to the energy being transferred to other harmonics. Figure 5 shows the comparison of the pressure-dependent peak attenuation estimated from Medwin's formula and Eq.(24). In this figure, the attenuation coefficient estimated by the linear theory is constant in the pressure range of $0.1p_0$ to $p_0$, while the attenuation coefficient estimated by the second-order theory shows a decreasing trend with increasing the excitation pressure. This discrepancy between the experimental and theoretical results lies in the nonlinear behaviors of shell parameters. The phenomenon of increased peak attenuation of lipid-coated UCAs was interpreted using a phenomenological nonlinear relation of shell rheology in our previous study (Xia et al., 2015). However, here the phenomenon of decreased peak attenuation of UCA near the main resonance is caused by its nonlinear oscillations. In reality, these two nonlinear behaviors of an UCA could coexist and simultaneously contribute to the nonlinear attenuation phenomena.

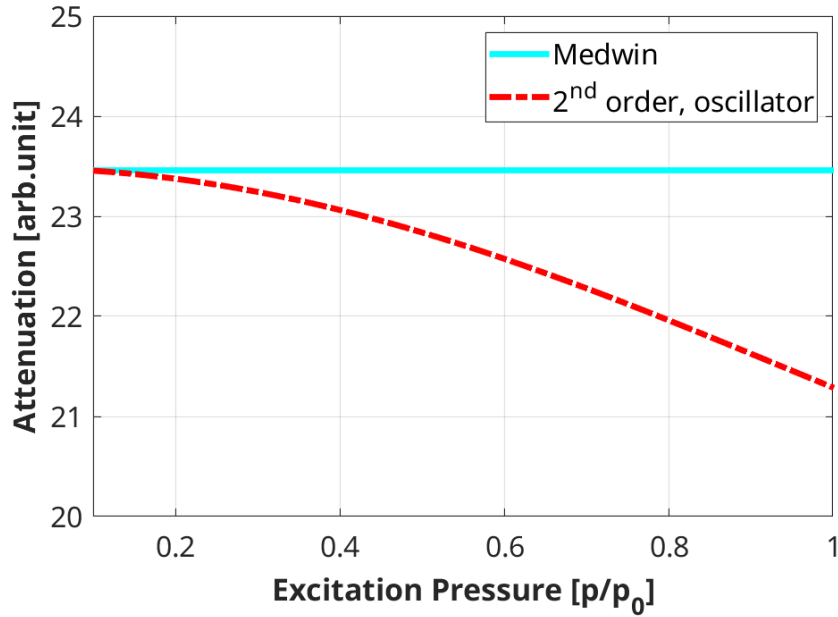

**Figure 5**: The pressure-dependent attenuation at the resonance frequency of an UCA.

### D. Comment on the effect of size distribution

By far we have only considered the monodisperse distribution of UCA population in the suspension in the investigation of attenuation. Although the polydisperse population of UCAs won't affect the result as long as the UCAs at small amplitude oscillations, the resonant or nonlinear oscillations still depend very much on the UCA sizes. For instance, in an acoustic field small bubbles tend to be 'stiffer' and exhibit less nonlinear oscillations, whereas big bubbles of the same shell parameters tend to be 'softer' and show more nonlinear oscillations. The coexistence of linear and nonlinear oscillations in an UCA suspension could smear out or enhance the nonlinear attenuation phenomenon. To investigate the effect of size distribution, we employed a normal distribution function defined by the mean radius $R_m$ and the standard deviation parameter $\sigma$. In the computation, the mean radius was assumed to $R_m = R_0 = 2.5$ µm, and the standard deviation was varied between 0.05 and 1. The bin size of the distribution was 0.05 µm. Note that a larger $\sigma$ indicates a broader distribution. These UCAs, with the same shell parameters as precious analysis, were excited using a pressure of $0.8p_0$. The comparison of the attenuation coefficient estimated by Medwin's formula Eq.(12) and the proposed formula Eq.(24) is presented in Figure 6. For $\sigma$ near 0.05, where the UCA population is almost monodisperse, the differences between Medwin's and $2^{nd}$ order results are restricted in the region of $f/f_0 =1$. Here $f_0$ is the resonance frequency of UCAs with the mean radius. In other words, the inaccuracy of the linear result only occurs near the resonance frequency provided that the UCA population is monodisperse. However, when $\sigma$ increases (the UCA population becomes more polydisperse), the difference of the attenuation curve spread out the resonant region, indicating that more discrepancies are introduced by the linear attenuation theory. Therefore, the proposed formula Eq.(24), well predicted the attenuation of monodisperse UCAs, has even better accuracy than Medwin's formula in the estimation of the attenuation coefficient of UCA suspensions involving polydisperse size distribution.

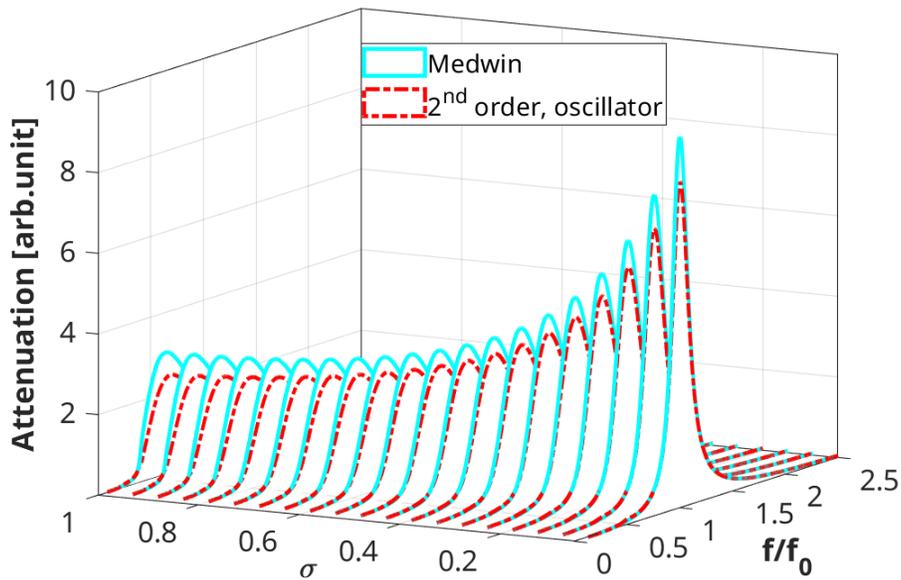

**Figure 6**: The frequency-dependent attenuation curves *vs.* different size distribution $\sigma$.

## V. CONCLUSION

By virtue of the second-order perturbative approximations, we have obtained formulae capable of evaluating the attenuation coefficient of an UCA (or UCA suspension) at resonant oscillations, extended the classic linear attenuation formula to accounting for nonlinear oscillations at the resonance frequency. These results can be readily adapted to a variety of models for the surface encapsulating of UCAs or free bubbles. Simulated results for an UCA with constant shell parameters at the excitation pressure of $0.8p_0$ showed the peak attenuation coefficient near the main resonance is 13% less than that estimated from the Medwin's formula, suggesting the improvement is prominent, especially for UCA suspensions with polydisperse population. The attenuation of resonant UCAs estimated by the proposed formulae showed excellent agreements with the results obtained from the full wave propagation equation. These results also demonstrated that the dynamic behaviors of the UCA at the resonance were nonlinear regardless of the employed low excitation pressures. Thus, the proposed formulae Eq.(24) and Eq.(36) have better accuracy than the linear theory [Eq.(12)] in predicting the attenuation coefficient of the resonant UCA. Furthermore, the attenuation coefficient of the UCA at the main resonance showed an inverse relationship to the excitation pressures, different from the experimental observations, indicating that the attenuation phenomenon involved in UCA oscillations is rather complicated. Both the nonlinear dynamic behaviors and shell properties could conspire to trigger the nonlinear attenuation behaviors. Therefore, further exploration of the nonlinear attenuation phenomenon of UCAs can focus on incorporating the effects of the nonlinear shell behaviors and nonlinear oscillations simultaneously into an attenuation theory.

**REFERENCES**

Ainslie, M. A., & Leighton, T. G. (2009). Near resonant bubble acoustic cross-section corrections, including examples from oceanography, volcanology, and biomedical ultrasound. *The Journal of the Acoustical Society of America, 126*(5), 2163-2175.

Ainslie, M. A., & Leighton, T. G. (2011). Review of scattering and extinction cross-sections, damping factors, and resonance frequencies of a spherical gas bubble. *The Journal of the Acoustical Society of America, 130*(5), 3184-3208.

Chatterjee, D., & Sarkar, K. (2003). A Newtonian rheological model for the interface of microbubble contrast agents. *Ultrasound in medicine & biology, 29*(12), 1749-1757.

Chen, Q., Zagzebski, J., Wilson, T., & Stiles, T. (2002). Pressure-dependent attenuation in ultrasound contrast agents. *Ultrasound in medicine & biology, 28*(8), 1041-1051.

Church, C. C. (1995). The effects of an elastic solid surface layer on the radial pulsations of gas bubbles. *The Journal of the Acoustical Society of America, 97*(3), 1510-1521.

Commander, K. W., & Prosperetti, A. (1989). Linear pressure waves in bubbly liquids: Comparison between theory and experiments. *The Journal of the Acoustical Society of America, 85*(2), 732-746.

De Jong, N., Emmer, M., Van Wamel, A., & Versluis, M. (2009). Ultrasonic characterization of ultrasound contrast agents. *Medical & biological engineering & computing, 47*(8), 861-873.

de Jong, N., Frinking, P. J., Bouakaz, A., Goorden, M., Schourmans, T., Jingping, X., & Mastik, F. (2000). Optical imaging of contrast agent microbubbles in an ultrasound field with a 100-MHz camera. *Ultrasound in medicine & biology, 26*(3), 487-492.

de Jong, N., Hoff, L., Skotland, T., & Bom, N. (1992). Absorption and scatter of encapsulated gas filled microspheres: theoretical considerations and some measurements. *Ultrasonics, 30*(2), 95-103.

Doinikov, A. A., & Bouakaz, A. (2011). Review of shell models for contrast agent microbubbles. *IEEE Transactions on Ultrasonics, Ferroelectrics, and Frequency Control, 58*(5), 981-993.

Doinikov, A. A., Haac, J. F., & Dayton, P. A. (2009). Resonance frequencies of lipid-shelled microbubbles in the regime of nonlinear oscillations. *Ultrasonics, 49*(2), 263-268.

Emmer, M., Vos, H. J., Goertz, D. E., van Wamel, A., Versluis, M., & de Jong, N. (2009). Pressure-dependent attenuation and scattering of phospholipid-coated microbubbles at low acoustic pressures. *Ultrasound in medicine & biology, 35*(1), 102-111.

Faez, T., Emmer, M., Kooiman, K., Versluis, M., van der Steen, A. F., & de Jong, N. (2013). 20 years of ultrasound contrast agent modeling. *Ultrasonics, Ferroelectrics, and Frequency Control, IEEE Transactions on, 60*(1).

Feng, Z., & Leal, L. (1993). On energy transfer in resonant bubble oscillations. *Physics of Fluids A: Fluid Dynamics, 5*(4), 826-836.

Gong, Y., Cabodi, M., & Porter, T. M. (2014). Acoustic investigation of pressure-dependent resonance and shell elasticity of lipid-coated monodisperse microbubbles. *Applied Physics Letters, 104*(7), 074103.

Helfield, B. L., & Goertz, D. E. (2013). Nonlinear resonance behavior and linear shell estimates for Definity™ and MicroMarker™ assessed with acoustic microbubble spectroscopy. *The Journal of the Acoustical Society of America, 133*(2), 1158-1168.

Hilgenfeldt, S., Lohse, D., & Zomack, M. (1998). Response of bubbles to diagnostic ultrasound: a unifying theoretical approach. *The European Physical Journal B-Condensed Matter and Complex Systems, 4*(2), 247-255.

Hoff, L. (2001). *Acoustic characterization of contrast agents for medical ultrasound imaging*: Springer Science & Business Media.

Hoff, L., Sontum, P. C., & Hovem, J. M. (2000). Oscillations of polymeric microbubbles: Effect of the encapsulating shell. *The Journal of the Acoustical Society of America, 107*(4), 2272-2280.

Karandish, F., Mamnoon, B., Feng, L., Haldar, M. K., Xia, L., Gange, K. N., . . . Mallik, S. (2018). Nucleus-Targeted, Echogenic Polymersomes for Delivering a Cancer Stemness Inhibitor to Pancreatic Cancer Cells. *Biomacromolecules, 19*(10), 4122-4132.

Khismatullin, D. B. (2004). Resonance frequency of microbubbles: Effect of viscosity. *The Journal of the Acoustical Society of America, 116*(3), 1463-1473.

Kulkarni, P., Haldar, M. K., Karandish, F., Confeld, M., Hossain, R., Borowicz, P., . . . Mallik, S. (2018). Tissue‐Penetrating, Hypoxia‐Responsive Echogenic Polymersomes For Drug Delivery To Solid Tumors. *Chemistry–A European Journal, 24*(48), 12490-12494.

Lauterborn, W., & Kurz, T. (2010). Physics of bubble oscillations. *Reports on progress in physics, 73*(10), 106501.

Marmottant, P., van der Meer, S., Emmer, M., Versluis, M., de Jong, N., Hilgenfeldt, S., & Lohse, D. (2005). A model for large amplitude oscillations of coated bubbles accounting for buckling and rupture. *The Journal of the Acoustical Society of America, 118*(6), 3499-3505.

Medwin, H. (1977). Counting bubbles acoustically: a review. *Ultrasonics, 15*(1), 7-13.

Overvelde, M., Garbin, V., Sijl, J., Dollet, B., De Jong, N., Lohse, D., & Versluis, M. (2010). Nonlinear shell behavior of phospholipid-coated microbubbles. *Ultrasound in medicine & biology, 36*(12), 2080-2092.

Paul, S., Katiyar, A., Sarkar, K., Chatterjee, D., Shi, W. T., & Forsberg, F. (2010). Material characterization of the encapsulation of an ultrasound contrast microbubble and its subharmonic response: Strain-softening interfacial elasticity model. *The Journal of the Acoustical Society of America, 127*(6), 3846-3857.

Prosperetti, A. (1977). Thermal effects and damping mechanisms in the forced radial oscillations of gas bubbles in liquids. *The Journal of the Acoustical Society of America, 61*(1), 17-27.

Prosperetti, A. (2013). A general derivation of the subharmonic threshold for non-linear bubble oscillations. *The Journal of the Acoustical Society of America, 133*(6), 3719-3726.

Sarkar, K., Shi, W. T., Chatterjee, D., & Forsberg, F. (2005). Characterization of ultrasound contrast microbubbles using in vitro experiments and viscous and viscoelastic interface models for encapsulation. *The Journal of the Acoustical Society of America, 118*(1), 539-550.

Segers, T., de Jong, N., & Versluis, M. (2016). Uniform scattering and attenuation of acoustically sorted ultrasound contrast agents: Modeling and experiments. *The Journal of the Acoustical Society of America, 140*(4), 2506-2517.

Tang, M.-X., & Eckersley, R. J. (2007). Frequency and pressure dependent attenuation and scattering by microbubbles. *Ultrasound in medicine & biology, 33*(1), 164-168.

Tang, M.-X., Eckersley, R. J., & Noble, J. A. (2005). Pressure-dependent attenuation with microbubbles at low mechanical index. *Ultrasound in medicine & biology, 31*(3), 377-384.

Tu, J., Swalwell, J. E., Giraud, D., Cui, W., Chen, W., & Matula, T. J. (2011). Microbubble sizing and shell characterization using flow cytometry. *IEEE Transactions on Ultrasonics, Ferroelectrics, and Frequency Control, 58*(5), 955-963.

Versluis, M., Goertz, D. E., Palanchon, P., Heitman, I. L., van der Meer, S. M., Dollet, B., . . . Lohse, D. (2010). Microbubble shape oscillations excited through ultrasonic parametric driving. *Physical Review E, 82*(2), 026321.

Xia, L. (2018). *Dynamics of Ultrasound Contrast Agents and Nonlinear Acoustic Waves: Experiments, Modeling, and Theories.* The George Washington University,

Xia, L. (2019). Analysis of acoustic nonlinearity parameter B/A in liquids containing ultrasound contrast agents. *The Journal of the Acoustical Society of America, 146*(2), 1394-1403. doi:10.1121/1.5123486

Xia, L., Karandish, F., Kumar, K. N., Froberg, J., Kulkarni, P., Gange, K. N., . . . Sarkar, K. (2017). Acoustic Characterization of Echogenic Polymersomes Prepared From Amphiphilic Block Copolymers. *Ultrasound in medicine & biology, 44*(2), 447-457.

Xia, L., Porter, T. M., & Sarkar, K. (2015). Interpreting attenuation at different excitation amplitudes to estimate strain-dependent interfacial rheological properties of lipid-coated monodisperse microbubbles. *The Journal of the Acoustical Society of America, 138*(6), 3994-4003.

Zhang, Y. (2013). A generalized equation for scattering cross section of spherical gas bubbles oscillating in liquids under acoustic excitation. *Journal of fluids engineering, 135*(9).

Versluis, M., Goertz, D. E., Palanchon, P., Heitman, I. L., van der Meer, S. M., Dollet, B., . . . Lohse, D. (2010). Microbubble shape oscillations excited through ultrasonic parametric driving. *Physical Review E, 82*(2), 026321.

Xia, L. (2018). *Dynamics of Ultrasound Contrast Agents and Nonlinear Acoustic Waves: Experiments, Modeling, and Theories.* The George Washington University,

Xia, L. (2019). Analysis of acoustic nonlinearity parameter B/A in liquids containing ultrasound contrast agents. *The Journal of the Acoustical Society of America, 146*(2), 1394-1403. doi:10.1121/1.5123486

Xia, L., Karandish, F., Kumar, K. N., Froberg, J., Kulkarni, P., Gange, K. N., . . . Sarkar, K. (2017). Acoustic Characterization of Echogenic Polymersomes Prepared From Amphiphilic Block Copolymers. *Ultrasound in medicine & biology, 44*(2), 447-457.

Xia, L., Porter, T. M., & Sarkar, K. (2015). Interpreting attenuation at different excitation amplitudes to estimate strain-dependent interfacial rheological properties of lipid-coated monodisperse microbubbles. *The Journal of the Acoustical Society of America, 138*(6), 3994-4003.

Zhang, Y. (2013). A generalized equation for scattering cross section of spherical gas bubbles oscillating in liquids under acoustic excitation. *Journal of fluids engineering, 135*(9).